\documentclass[prl,twocolumn,groupedaddress,showpacs,nofootinbib]{revtex4}
\usepackage{graphicx}
\usepackage{dcolumn}% Align table columns on decimal point
\usepackage{bm}% bold math
\usepackage{amssymb}
\usepackage{mathrsfs}
\usepackage{amsmath}
\usepackage{epsfig}
\usepackage{here}
\usepackage[dvips]{color}
\usepackage{hhline}

\begin{document}

\title{Annihilating Leptogenesis}

\author{Pei-Hong Gu$^{1}_{}$}
\email{pgu@ictp.it}

\author{Utpal Sarkar$^{2,3}_{}$}
\email{utpal@prl.res.in}

\affiliation{$^{1}_{}$The Abdus Salam International Centre for
Theoretical Physics, Strada Costiera 11, 34014 Trieste, Italy\\
$^{2}_{}$Physical Research Laboratory, Ahmedabad 380009, India\\
$^3$Department of Physics and McDonnell Center for the Space
Sciences, Washington University, St. Louis, MO 63130, USA}

\begin{abstract}

Leptogenesis is usually realized through decays of heavy particles.
In this article we consider another possibility of generating a
lepton asymmetry through annihilations of heavy particles. We
demonstrate our idea with a realistic extension of the standard
model containing a heavy doublet and a light singlet scalars in
addition to right-handed neutrinos and Higgs triplets required for
type-I+II seesaw of neutrino masses. We also clarify that this
annihilating leptogenesis scenario can be naturally embedded in more
fundamental theories, like left-right symmetric models or grand
unified theories.

\end{abstract}


\maketitle

\vspace{5mm}

\emph{Introduction}: It is well known that any CPT-conserved
baryogenesis theories should satisfy Sakharov conditions
\cite{sakharov1967}: (1) baryon number violation, (2) C and CP
nonconservation, (3) thermodynamic nonequilibrium. In the standard
model (SM), there is a $\textrm{SU(2)}_L$ global anomaly
\cite{thooft1976}, which causes baryon number $(B)$ and lepton
number $(L)$ violation, although their difference $B-L$ is still
conserved. This anomaly induced $B+L$ violating processes will be
suppressed by quantum tunneling probability at zero temperature,
however, at finite temperature it will become fast in presence of an
instanton-like solution, the sphaleron \cite{krs1985}. The sphaleron
action can partially convert an existing $B-L$ asymmetry to a baryon
asymmetry and then we can understand the matter-antimatter asymmetry
of the present Universe. With this essence, Fukugita and Yanagida
proposed the leptogenesis \cite{fy1986} scenario, where a lepton
asymmetry (without baryon asymmetry) is generated before the
sphaleron processes become very slow around the time of the
electroweak phase transition. In the conventional leptogenesis
scenario, the lepton asymmetry is produced by decays of heavy
particles \cite{fy1986,lpy1986,crv1996}. We argue here that it is
unnecessary to constrain the leptogenesis mechanism in the decaying
scenario, since the required conditions: lepton number violation, C
and CP violation, and departure from equilibrium, can also be
reached in annihilating processes. In the following, we demonstrate
the possibility of "annihilating leptogenesis" in a realistic model.

\vspace{5mm}

\emph{The model}: For simplicity, we only write down the following
mass terms and interactions that are relevant for the rest of our
discussions,
\begin{widetext}
\begin{eqnarray}
\label{lagrangian}
\mathcal{L}&\supset&\left[-\frac{1}{2}M_{N}^{}\overline{N_R^c}N_R^{}
-y\overline{\psi_{L}^{}}\phi N_{R}^{}+\textrm{H.c.}
\right]-M_\xi^2\textrm{Tr}\left(\xi^\dagger_{}\xi\right)
-\left[\frac{1}{2}f\overline{\psi_{L}^{c}} i\tau_{2}
\xi\psi_{L}+\mu\phi^{T}i\tau_{2}\xi\phi
+\textrm{H.c.}\right]\nonumber\\[3mm]
&&-M_\eta^{2}\eta^{\dagger}_{}\eta -
\left[\mu'\eta^{T}_{}i\tau_{2}^{}\xi\eta
\phantom{\frac{a}{b}}+\lambda (\phi^\dagger \eta)^2_{}+\textrm{H.c.}\right]
-\left(\rho\chi\eta^\dagger_{}\phi-\kappa\chi\eta^T_{}i\tau_2^{}\xi\phi+\textrm{H.c.}\right)\nonumber \\[3mm]
&& - \alpha_1^{} \left(\eta^\dagger \eta\right)\left(\phi^\dagger
\phi\right) -\alpha_2^{} \left(\eta^\dagger \phi\right)
\left(\phi^\dagger \eta\right) -\alpha_3^{}
 \chi^2_{}\left(\eta^\dagger \eta\right)-\alpha_{4}^{}
 \left(\eta^\dagger \eta\right)\textrm{Tr}\left(\xi^\dagger\xi\right)\,.
\end{eqnarray}
\end{widetext} Here $\psi_{L}^{}$ and $\phi$, respectively, are the
SM lepton and Higgs doublets, $N_{R}^{}$ denotes the right-handed
neutrinos, $\xi$ stands for the Higgs triplets, $\eta$ is a heavy
doublet scalar carrying the same quantum number with $\phi$, and
$\chi$ is a light singlet scalar. We impose a $Z_2^{}$ discrete
symmetry, under which only $\eta$ and $\chi$ are odd-parity fields
and hence are protected from nonzero vacuum expectation value (VEV).

Corresponding to $L=+1$ for the SM leptons and $L=0$ for the SM
Higgs doublet, we assign $L=+1$ for the right-handed neutrinos and
$L=-2$ for the Higgs triplets. Therefore, the lepton number is
explicitly violated by the Majorana masses of the right-handed
neutrinos, the trilinear interactions between the triplet and
doublet scalars and the quartic interactions among the triplet,
doublet and singlet scalars. We consider this as an effective theory
from a more fundamental theory, where the lepton number is
spontaneously broken locally or globally. As a result of the lepton
number violation, the type-I and II seesaw mechanisms
\cite{minkowski1977,mw1980} can be realized phenomenologically. The
resulting neutrino masses would be
\begin{eqnarray}
\mathcal{L}_{mass}^{\nu}=\frac{1}{2}m_{\nu}^{}\overline{\nu_{L}^{c}}\nu_{L}^{}+\textrm{h.c.}\,,
\end{eqnarray}
where the mass matrix $m_{\nu}^{}$ contains two parts,
\begin{eqnarray}
m_{\nu}^{}=m_{\nu}^{\textrm{I}}+m_{\nu}^{\textrm{II}}\,,
\end{eqnarray}
with the type-I seesaw,
\begin{eqnarray}
m_{\nu}^{\textrm{I}}=-y^\ast_{}\frac{v^{2}_{}}{M_{N}^{}}y^\dagger_{}\,,
\end{eqnarray}
and the type-II seesaw,
\begin{eqnarray}
m_{\nu}^{\textrm{II}}=-f\frac{\mu^\ast_{}v^{2}_{}}{M_{\xi}^{2}}\,.
\end{eqnarray}
Here $v\equiv \langle\phi\rangle\simeq 174\,\textrm{GeV}$ is the VEV
of the SM Higgs doublet.

\vspace{5mm}

\begin{figure*}
\vspace{6.0cm} \epsfig{file=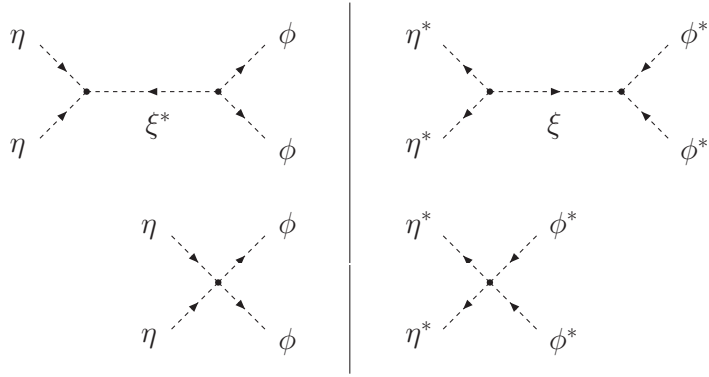, bbllx=6.3cm, bblly=6.0cm,
bburx=16.3cm, bbury=16cm, width=8cm, height=8cm, angle=0, clip=0}
\vspace{-8cm} \caption{\label{scalar} The heavy doublet scalar
annihilates into the SM Higgs doublet at tree level. }
\end{figure*}

\begin{figure*}
\vspace{8.5cm} \epsfig{file=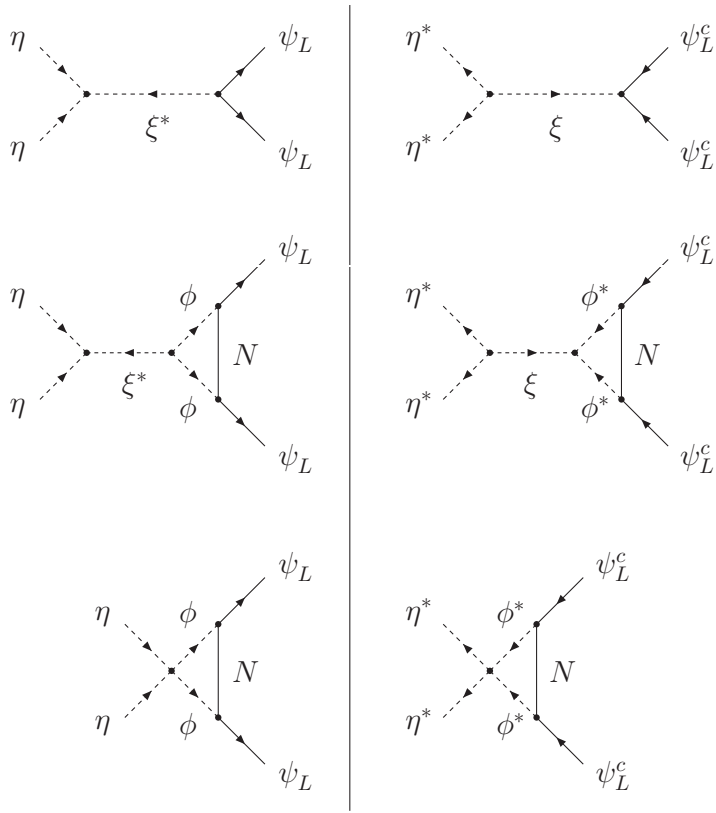, bbllx=6.3cm, bblly=6.0cm,
bburx=16.3cm, bbury=16cm, width=8cm, height=8cm, angle=0, clip=0}
\vspace{-2.0cm} \caption{\label{fermion} The heavy doublet scalar
annihilates into the SM lepton doublet at tree level and one-loop
order. Here we don't consider the cases with more than two Higgs
triplets, which can result in self-energy correction even if the
right-handed neutrinos are absent. }
\end{figure*}

\emph{CP-asymmetry}: The heavy doublet scalar can annihilate into
two SM Higgs or lepton doublets as shown in Figs. \ref{scalar} and
\ref{fermion}. At the tree level, the cross sections for the
CP-conjugate channels, $(\eta\eta \rightarrow \phi\phi,\,\,
\eta^\ast_{}\eta^\ast_{} \rightarrow \phi_{}^{\ast}\phi_{}^{\ast})$
and $(\eta\eta \rightarrow \psi_{L}^{}\psi_{L}^{},\,\,
\eta^\ast_{}\eta^\ast_{} \rightarrow \psi_{L}^{c}\psi_{L}^{c})$ are
given by
\begin{subequations}
\begin{eqnarray}
&& \label{cross-section-1}\sigma_{ \eta\eta \rightarrow \phi\phi}^{}
|\vec{v}|=\sigma_{ \eta^\ast_{}\eta^\ast_{} \rightarrow
\phi_{}^{\ast}\phi_{}^{\ast}}^{} |\vec{v}|\nonumber\\
&=&\frac{1}{4\pi}\left|\lambda-\frac{\mu'\mu^\ast_{}}{M_\xi^2}\right|^{2}_{}\frac{1}{s}\,,\\
&& \label{cross-section-2}\sigma_{ \eta\eta \rightarrow
\psi_{L}^{}\psi_{L}^{}}^{} |\vec{v}|=\sigma_{
\eta^\ast_{}\eta^\ast_{} \rightarrow
\psi_{L}^{c}\psi_{L}^{c}}^{} |\vec{v}|\nonumber\\
&=&\frac{1}{4\pi}\frac{|\mu'|^2_{}}{M_\xi^4}\textrm{Tr}\left(f^\dagger_{}f\right)\,.
\end{eqnarray}
\end{subequations}
Here $s\geq 4 M_\chi^2$ is the squared center of mass energy and
$|\vec{v}|=2\left(1-4M_{\chi}^{2}/s\right)^{\frac{1}{2}}_{}$ is the
relative velocity between the two heavy doublets. Note in the above
and following calculations the Higgs triplets are assumed to be much
heavier than the heavy doublet.

At the one-loop order, the vertex correction induced by the
right-handed neutrinos and/or the self-energy correction induced by
the Higgs triplets (if there are more than two Higgs triplets), will
result in a difference between the cross sections of the
CP-conjugate channels, as long as the CP is not conserved. For
example, we consider the case only with one Higgs triplet and then
derive the CP-asymmetry,
\begin{eqnarray}
\varepsilon&\equiv&2\frac{\sigma_{ \eta\eta \rightarrow
\psi_{L}^{}\psi_{L}^{}}^{}-\sigma_{ \eta^\ast_{}\eta^\ast_{}
\rightarrow \psi_{L}^{c}\psi_{L}^{c}}}{\sigma_{ \eta\eta \rightarrow
\psi_{L}^{}\psi_{L}^{}}^{}+\sigma_{ \eta\eta \rightarrow
\phi\phi}^{}}\nonumber\\
&\equiv&2\frac{\sigma_{ \eta\eta \rightarrow
\psi_{L}^{}\psi_{L}^{}}^{}-\sigma_{ \eta^\ast_{}\eta^\ast_{}
\rightarrow \psi_{L}^{c}\psi_{L}^{c}}}{\sigma_{
\eta^\ast_{}\eta^\ast_{} \rightarrow
\psi_{L}^{c}\psi_{L}^{c}}^{}+\sigma_{ \eta^\ast_{}\eta^\ast_{}
\rightarrow
\phi^\ast_{}\phi^\ast_{}}^{}}\nonumber\\
&\simeq&\frac{1}{4\pi}\frac{\textrm{Im}\left[\textrm{Tr}\left(m_\nu^{\textrm{I}\dagger}
m_\nu^\textrm{II}\right)\right]}{\textrm{Tr}\left(m_\nu^{\textrm{II}\dagger}
m_\nu^\textrm{II}\right)}\frac{|\mu|^{2}_{}}{M_{\xi}^{2}}\textrm{Br}\,,
\end{eqnarray}
where the branch ratio is defined as
\begin{eqnarray}
\textrm{Br}&=&\frac{\sigma_{ \eta\eta \rightarrow
\psi_{L}^{}\psi_{L}^{}}^{}}{\sigma_{ \eta\eta \rightarrow
\psi_{L}^{}\psi_{L}^{}}^{}+\sigma_{ \eta\eta \rightarrow
\phi^{}_{}\phi^{}_{}}^{}}\nonumber\\
&=&\frac{\textrm{Tr}\left(f^\dagger_{}f\right)}{\textrm{Tr}\left(f^\dagger_{}f\right)+\left|\lambda
M_\xi^2/\mu'-\mu^\ast_{}\right|^{2}_{}/s}\,.
\end{eqnarray}

\vspace{5mm}

\emph{Frozen Temperature}: The heavy doublet scalar have four types
of interactions, including (1) the gauge couplings; (2) the quartic
couplings with other scalars; (3) the trilinear couplings with the
triplet Higgs scalar; (4) the trilinear couplings with the light
singlet scalar and the doublet Higgs scalar. The induced
annihilating and decaying processes will determine the frozen
temperature $T_{F}^{}$, below which the heavy doublet will deviate
from its equilibrium distribution. We now analyze these processes.
Firstly, the gauge interactions can be safely ignored when the heavy
doublet is very heavy (roughly
$\gtrsim\mathcal{O}(10^{10}_{}\,\textrm{GeV})$). Secondly, we assume
the decays of the heavy doublet has been decoupled before it became
non-relativistic. This yields
\begin{eqnarray}
\label{oec1} \Gamma_{D}^{}\ll
H(T)\left|_{T=M_{\eta}^{}}^{}\right.\,,
\end{eqnarray}
where
\begin{eqnarray}
\Gamma_{D}^{}&=&\Gamma(\eta\rightarrow\chi\phi)=\Gamma(\eta^\ast_{}\rightarrow\chi\phi^\ast_{})\nonumber\\
&=&\frac{1}{16\pi}\frac{|\rho|^{2}_{}}{M_{\eta}^{}}\,,
\end{eqnarray}
is the decay width \footnote{The light singlet scalar $\chi$ is
stable and will contribute a relic density to the Universe. In the
presence of its quartic coupling to the SM Higgs doublet $\phi$,
i.e.
$\mathcal{L}\supset\epsilon\chi^2_{}\left(\phi^\dagger_{}\phi\right)+\textrm{H.c.}$,
its frozen temperature and then its relic density can be determined
by its annihilations into the SM fields, including the quarks, the
leptons, the gauge bosons, and the Higgs boson. For appropriate
choice of the coefficient $\epsilon$, this singlet scalar $\chi$
with a mass of the order of $\mathcal{O}(\textrm{GeV} -
\textrm{TeV})$ can leave a desired relic density to serve as the
dark matter \cite{sz1985}.} and
\begin{eqnarray}
\label{hubble}
H(T)&=&\left(\frac{8\pi^{3}_{}g_{\ast}^{}}{90}\right)^{\frac{1}{2}}_{}\frac{T^{2}_{}}{M_{\textrm{Pl}}^{}}\,
\end{eqnarray}
is the Hubble constant with $M_{\textrm{Pl}}^{}\simeq
10^{19}_{}\,\textrm{GeV}$ and $g_{\ast}^{}=\mathcal{O}(100)$. Here
the four-body decays mediated by the imaginary Higgs triplet are
greatly suppressed and hence have been ignored in the decay width.
The condition (\ref{oec1}) is easy to realize. For example, we input
$M_{\eta}^{}=10^{12}_{}\,\textrm{GeV}$ and
$|\rho|=10^{9}_{}\,\textrm{GeV}$. Thirdly, we consider the
annihilating processes shown in Figs. \ref{scalar} and
\ref{fermion}. Their frozen temperature can be solved through the
following out-of-equilibrium condition,
\begin{eqnarray}
\label{oec2} \Gamma_{A}^{}\simeq H(T)\,,
\end{eqnarray}
where
\begin{eqnarray}
\Gamma_{A}^{}=n_{\eta}^{\textrm{eq}}\left\langle\left(\sigma_{\eta\eta\rightarrow
\psi_{L}^{}\psi_{L}^{}}^{} + \sigma_{\eta\eta\rightarrow
\phi\phi}^{}\right ) |\vec{v}|\right\rangle  \,,
\end{eqnarray}
is the rate with
\begin{eqnarray}
n_{\eta}^{\textrm{eq}}=
\frac{2}{\pi^2}T^{3}_{}\frac{M_{\eta}^{2}}{T^{2}_{}}K_2^{}\left(\frac{M_{\eta}^{}}{T}\right)
\end{eqnarray}
being the equilibrium distribution of number density. Finally, we
study other annihilating processes. For $M_{\eta}^{}\ll M_{\xi}^{}$,
only the $2\rightarrow 2$ processes
$(\eta\eta^\ast_{}\rightarrow\phi\phi^\ast_{},\chi\chi)$ are not
suppressed. We can replace
$|\lambda-\mu'\mu^\ast_{}/M_\xi^2|^{2}_{}$ by
$\alpha_{1}^{2}+\alpha_{2}^{2}+\alpha_{1}^{}\alpha_{2}^{}+2\alpha_{3}^{2}$
in Eq. (\ref{cross-section-1}) and then get the cross section. In
the following section, we will take
$\alpha_{1}^{2}+\alpha_{2}^{2}+\alpha_{1}^{}\alpha_{2}^{}+2\alpha_{3}^{2}\ll|\lambda-\mu'\mu^\ast_{}/M_\xi^2|^{2}_{}$
to simply give a frozen temperature for discussing the final baryon
asymmetry.

\vspace{5mm}

\emph{Final Baryon Asymmetry}: The right-handed neutrinos and/or
Higgs triplets in the seesaw context will mediate some lepton number
violating processes, including
\begin{eqnarray}
\label{delta2} \phi\phi\leftrightarrow\psi_{L}^{}\psi_{L}^{}\,,~
\phi^\ast_{}\phi^\ast_{}\leftrightarrow\psi_{L}^{c}\psi_{L}^{c}\,,~\textrm{and}~
\psi_{L}^{c}\phi\leftrightarrow\psi_{L}^{}\phi_{}^{\ast}\,.
\end{eqnarray}
These processes will erase any lepton asymmetry produced before
their departure from equilibrium. At low temperatures (compared with
the masses of the right-handed neutrinos and/or Higgs triplets), the
processes (\ref{delta2}) will take place with the rate
\cite{fy1990},
\begin{eqnarray}
\Gamma_{\Delta L
=2}^{}=\frac{1}{\pi^{3}_{}}\frac{T^{3}_{}}{v^{4}_{}}\textrm{Tr}\left(m_{\nu}^{\dagger}m_{\nu}^{}\right)\,.
\end{eqnarray}
Requiring the above rate to be smaller than the Hubble constant, one can
yield a decoupled temperature,
\begin{eqnarray}
T_{D}^{}=
10^{12}_{}\,\textrm{GeV}\,\left[\frac{\left(0.2\,\textrm{eV}\right)^{2}_{}}{\textrm{Tr}\left(m_{\nu}^{\dagger}m_{\nu}^{}\right)}\right]\,,
\end{eqnarray}
above which any existing lepton asymmetry will be washed out. By
inputting the neutrino masses from the neutrino oscillation
experiments and cosmological observations, it is straightforward to
see that $T_{D}^{} = \mathcal{O}(10^{12}\,\textrm{GeV})$. This also
means that for a typically seesaw scale
$\sim\mathcal{O}(10^{14}_{}\,\textrm{GeV})$, the decays of the
right-handed neutrinos and/or Higgs triplets will fail in generating
a desired lepton asymmetry.

We now demonstrate how the annihilating leptogenesis can be realized
in the present model. Here we do not attempt to solve the completed
Boltzmann equations, which will not give any better insight to the
problem and will be studied elsewhere. For simplicity, we consider
the case where $T_{D}^{}>T_{F}^{}$ with $T_{F}^{}$ being determined
by Eq. (\ref{oec2}) and hence the final baryon asymmetry can be well
described by
\begin{eqnarray}
\frac{n_{B}^{}}{s}=\frac{28}{79}\frac{n_{B-L}^{}}{s}
=-\frac{28}{79}\frac{n_{L}^{}}{s}
\simeq-\frac{28}{79}\left.\left[\langle\varepsilon\rangle\frac{n_{\eta}^{\textrm{eq}}}{s}\right]\right|_{T_F^{}}\,.
\end{eqnarray}
Here $s$ is the entropy density given by a very good approximation,
\begin{eqnarray}
s=\frac{2 \pi^2_{}}{45}g_{\ast s}^{} T^3_{}\,
\end{eqnarray}
with $g_{\ast s}^{}\simeq g_{\ast}^{}=\mathcal{O}(100)$. We consider
here a representative choice of parameters $f=\mathcal{O}(1)$,
$y=\mathcal{O}(1)$ and $|\mu|\lesssim M_{\xi}^{}\sim
M_{N}^{}=\mathcal{O}(10^{14}_{}\,\textrm{GeV})$, we obtain the
desired neutrino masses, $m_{\nu}^{\textrm{I}}\sim
m_{\nu}^{\textrm{II}}\sim m_{\nu}^{}=
\mathcal{O}(0.01-1\,\textrm{eV})$. We further consider
$M_{\eta}^{}=\mathcal{O}(10^{12}_{}\,\textrm{GeV})$,
$\mu'=\mathcal{O}(10^{13}_{}\,\textrm{GeV})$,
$\rho=\mathcal{O}(10^{9}_{}\,\textrm{GeV})$, $\lambda\sim
\mu'\mu^\ast_{}/M_{\xi}^{2}\sim \lambda-
\mu'\mu^\ast_{}/M_{\xi}^{2}=\mathcal{O}(0.1)$, $\kappa
<\mathcal{O}(1)$, $\alpha_{1,2,3,4}^{} =\mathcal{O}(0.01)$ and hence
obtain $T_{F}^{}= \mathcal{O}(10^{11-12}_{}\,\textrm{GeV})$ and
$n_{\eta}^{\textrm{eq}}/s=\mathcal{O}(10^{-(3-4)}_{})$. The thermal
averaged CP-asymmetry then comes out to be
$|\langle\varepsilon\rangle|\lesssim 10^{-5}_{}$, so that we are
flexible enough to get the desired baryon asymmetry
\begin{eqnarray}
\frac{n_B}{s}=\mathcal{O}(10^{-10}_{})\,.
\end{eqnarray}

\vspace{5mm}

\emph{Left-Right Symmetric Extension}: This model can be naturally
embedded in a left-right symmetric model or any grand unified
theory. In a left-right symmetric model, besides the usual
bi-doublet Higgs scalar $\Phi \equiv (1,2,2,0)$ [under the
left-right symmetric gauge group $SU(3)_{c}^{}\times SU(2)_L^{}
\times SU(2)_R^{} \times U(1)_{B-L}^{}$], we introduce a second
bi-doublet Higgs scalar $\Psi \equiv (1,2,2,0)$ and a singlet scalar
$\chi \equiv (1,1,1,0)$, which are odd under the $Z_2$ symmetry. The
discrete symmetry protects $\Psi$ and $\chi$ from any VEV and also
restricts its couplings. We also introduce the triplet Higgs scalar
$\Delta_R \equiv (1,1,3,-2)$ (to break the left-right symmetry) and
its counterpart $\Delta_L \equiv (1,3,1,-2)$. Then the lepton number
violating interactions for the type-I+II seesaw and the annihilating
leptogenesis will emerge when $\Delta_R$ acquires a VEV
$\langle\Delta_R\rangle$ to break the left-right symmetry. If we
embed this left-symmetric model in any grand unified theory, the
scale of left-right symmetry breaking comes out to be
$\langle\Delta_R\rangle
> 10^{13}\,\textrm{GeV}$, so that our choice of the seesaw scale
$\sim \langle \Delta_R \rangle = {\cal O}(10^{14}\,\textrm{GeV})$ is
a very natural one.

\vspace{5mm}

\emph{Conclusion}: In this paper, we proposed and demonstrated that
the leptogenesis could be realized in the annihilating scenario with
the necessary conditions: lepton number violation, C and CP
violation, and departure from equilibrium. We gave a realistic model
where (1) the right-handed neutrinos and Higgs triplets required for
the seesaw failed in realizing the decaying leptogenesis; (2) but
the heavy doublet scalar succeeded in generating a lepton asymmetry
through its annihilating to the SM lepton or Higgs doublet. The
"annihilating leptogenesis" can be realized in more fundamental
theories, like left-right symmetric theories or grand unified
theories, where the seesaw scale considered in the present model
comes out from an analysis of the gauge coupling unification.

\vspace{5mm}

\textbf{Acknowledgement}: PHG thanks Borut Bajc and Goran
Senjanovi$\rm\acute{c}$ for helpful discussions. US thanks the
Department of Physics and the McDonnell Center for the Space
Sciences at Washington University in St. Louis for inviting him as
Clark Way Harrison visiting professor and thanks R. Cowsik and F.
Ferrer for discussions. We thank the anonymous referee to point out
the flaw of the overabundant relic density in the original model.

\end{document}